\documentclass[prd,amsmath,amssymb,aps,longbibliography,superscriptaddress,twocolumn,nofootinbib,10pt,showkeys]{revtex4-1}
\usepackage{textcase}
\usepackage{amssymb,amsmath,mathtools} 
\usepackage{url}
\usepackage{natbib}

\usepackage{dcolumn}
\usepackage{bm}
\usepackage[T1]{fontenc}
\usepackage[utf8]{inputenc}
\usepackage{booktabs}
\usepackage{multirow}
\usepackage{graphicx}
\usepackage{xcolor}
\usepackage{microtype} 
\usepackage{hyperref} 

\def\aap{Astron. Astrophys.}
\def\mnras{Mon. Not. Roy. Astron. Soc.}
\def\apjl{Astrophys. J. Lett.}

\def\apjs{Astrophys. J., Suppl. Ser.}

\def\pasa{Publications of the Astronomical Society of Australia}
\hypersetup{
  colorlinks=true,
  linkcolor=red,
  citecolor=blue,
  urlcolor=blue
}
\begin{document}

\title{Improved Identification of Strongly Lensed Gravitational Waves with Host Galaxy Locations}

\author{Tonghua Liu}
\affiliation{School of Physics and Optoelectronic Engineering, Yangtze University, Jingzhou, 434023, China;}

\author{Kai Liao}
\email{liaokai@whu.edu.cn;}
\affiliation{School of Physics and Technology, Wuhan University, Wuhan 430072, China;}

\begin{abstract}
We present a Bayesian framework that enhances the identification of strongly lensed gravitational waves (GWs) by incorporating informative positional priors from the Euclid galaxy lens catalog. The core of our method introduces a two-step reweighting scheme: first, gravitational wave parameter estimation is performed under a uniform sky prior; the resulting posterior is then used to reweight galaxy positions within the Euclid catalog, constructing an astrophysically informed positional prior.
Comparing this Euclid-informed prior against a uniform prior within our framework reveals distinct behaviors. While the posterior estimates of the intrinsic waveform parameters show little sensitivity to the prior change, the Bayes factor for lensing identification exhibits significant prior dependence. Crucially, for truly lensed event pairs, the Bayes factor systematically increases, whereas for unlensed pairs it decreases. This dual effect is vital for robust discrimination.
Our analysis demonstrates that this multi-messenger approach significantly improves the confidence of lensing searches. For lensed pairs, the method boosts the Bayes factor by an average factor of $\sim 10$, while effectively suppressing false positives for unlensed coincidences. This underscores the critical importance of prior specification and showcases the substantial gains achievable by synergizing gravitational-wave data with electromagnetic survey information.

\end{abstract}

\maketitle

\section{Introduction}
The strong gravitational lensing of gravitational waves (GWs) offers a unique observational probe into cosmology, galactic structure, and fundamental physics. When GWs are lensed by foreground massive objects such as galaxies or clusters, they produce multiple images, each characterized by distinct time delays and magnification ratios \citep{1996PhRvL..77.2875W,2005PhRvD..71j1301Y,2003ApJ...595.1039T,2017arXiv170204724D,2021PhRvD.103f4047E}. Although no such events have been conclusively confirmed in current LIGO-Virgo-KAGRA (LVK) observations \citep{2021ApJ...923...14A,2019ApJ...874L...2H,2021SoftX..1300658A,2024ApJ...970..191A,2020PhRvD.102h4031M,2020arXiv200712709D,2021ApJ...908...97L}, third-generation detectors like the Einstein Telescope (ET) are expected to detect a large population of lensed events \citep{2022MNRAS.509.3772Y,2018MNRAS.476.2220L,2025arXiv250312263A,2022ApJ...929....9X}. This prospect opens new avenues for precision cosmological measurements \citep{2019RPPh...82l6901O,2022ChPhL..39k9801L} and motivates a major goal for multi-messenger astronomy. The scientific potential is compelling: lensed GWs can achieve arc-second sky localization \citep{2020MNRAS.498.3395H,2020MNRAS.497..204Y}, enable high-resolution mapping of lens mass distributions \citep{2018PhRvD..98h3005L,2019A&A...627A.130D,2025MNRAS.536.2212P}, provide precision measurements of cosmological parameters \citep{2011MNRAS.415.2773S,2019NatSR...911608C,2017NatCo...8.1148L,2023PhRvL.130z1401J,2020PhRvD.101f4011H}, yield novel  method for detecting dark matter \citep{2023PhRvD.108j3529T,2025PhRvL.135k1402J}, and test general relativity \citep{2017PhRvL.118i1101C,2021PhRvD.103b4038G,2017PhRvL.118i1102F}.

In searches for such lensed events, the Bayes factor serves as a key statistic for determining whether two detected signals originate from the same lensed source \citep{2018arXiv180707062H,2019PASA...36...10T,2023PhRvD.107l3015L,2023MNRAS.526..682G,2025NatAs...9..916S}. However, conventional methods often underutilize available independent information, such as the positional priors of the lens galaxy provided by galaxy surveys like Euclid \citep{2022A&A...662A.112E,2025arXiv250315324E,2025A&A...697A...1E}. This omission can lead to systematic biases and information loss. Recent studies have further highlighted that neglecting selection effects, realistic lens population distributions, and informed time delay priors can make the Bayes factor highly sensitive to model assumptions, thereby compromising the reliability of detection claims \citep{2021ApJ...923...14A,2018arXiv180707062H,2025arXiv251015463H}.

To address this gap and enhance the identification of lensed gravitational-wave events, we introduce a Bayesian framework that systematically integrates positional information from the Euclid galaxy catalog. Our approach implements a two-step reweighting scheme: (1) we first perform GW parameter estimation under a uniform sky prior to obtain the source localization posterior; (2) we then reweight the Euclid catalog using this posterior to construct an astrophysically informed spatial prior for the lens. This idea closely follows the multi-messenger methodology advocated by \citet{2014ApJ...795...43F}, who demonstrated how electromagnetic galaxy catalogs can substantially improve GW inference. Unlike earlier catalog-overlay techniques, our framework quantitatively merges the GW likelihood with galaxy-survey constraints, thereby providing a robust statistical measure to distinguish lensed from unlensed event pairs and suppress false positives from random sky alignments.

This work aims to establish a more robust statistical foundation for strong lensing searches by directly incorporating survey-informed priors. We present a quantitative assessment of how this integration improves search sensitivity and addresses the documented prior-sensitivity of the Bayes factor \citep{2019PASA...36...10T}, capitalizing on the synergy between third-generation gravitational-wave detectors and deep multi-wavelength surveys.

This paper is organized as follows. Section \ref{sec2} presents our methodology, describing the simulation of the Euclid-like lens galaxy catalog, the construction of the informed positional prior, and the Bayesian framework for lensing identification. Section \ref{sec3} shows the key results, focusing on the comparison of Bayes factors under uniform and Euclid-informed priors and the robustness of intrinsic parameter estimation. Section \ref{sec4} summarizes the conclusions and discusses future extensions. Unless otherwise stated, we assume a flat $\Lambda$CDM cosmology with $H_0 = 70.0~\mathrm{km~s^{-1}~Mpc^{-1}}$, $\Omega_{\mathrm{M}}=0.3$, and $c = G = 1$.

\section{Methodology}\label{sec2}
This section describes the complete framework for our lensing identification approach, including the simulation of the Euclid lens galaxy catalog, the construction of the informed positional prior, and the Bayesian inference setup for computing lensing Bayes factors.

\subsection{Euclid Lens Galaxy Catalog Simulation}
To systematically investigate the gravitational lensing effects on GW events detected by the Einstein Telescope (ET),  we developed a numerical simulation to generate a representative mock catalog of lens galaxies that matches the depth and sky density expected from the Euclid survey. The simulation is centered on a specific sky location (we take this input value as an example to illustrate our workflow): Right Ascension $\alpha=1.375$ radians and Declination $\delta=-1.2108$ radians, corresponding to the injection parameters of the GW signals in our study, ensuring a consistent spatial reference frame for lensing analysis.

The simulated region spans 5 square degrees, chosen to match the typical sky localization uncertainty of ET (a few square degrees for high-signal-to-noise ratio events) \citep{2025arXiv250312263A}, covering the complete positional prior of a single ET detection to study correlations between GW signals and foreground mass distributions.

The sky region is discretized into a uniform grid of $20 \times 20 = 400$ cells, each measuring $100'' \times 100''$ (arcseconds), balancing spatial resolution and computational efficiency. We adopt a lens galaxy surface density of 15 galaxies per square degree, consistent with the expected range (10–20 galaxies per square degree) from Euclid's deep survey observations \citep{2025arXiv250315324E}. The probability $P_{\text{cell}}$ of a galaxy residing in any cell is calculated as the product of the surface density $\rho$ and the cell area $A_{\text{cell}}$:
\begin{equation}
    P_{\text{cell}} = \rho \times A_{\text{cell}} \approx 1.157 \times 10^{-2},
\end{equation}
where the cell area $A_{\text{cell}} \approx 7.716 \times 10^{-4} \, \text{deg}^2$ (derived from $100'' \times 100''$). A Monte Carlo method populates the grid: galaxy presence in each cell is determined by $P_{\text{cell}}$, and precise positions are assigned uniformly within occupied cells to ensure a statistically authentic spatial distribution. This results in approximately 1.2\% chance per cell, leading to a realistic stochastic distribution of galaxies across the field.

The 5 square degree field yields an expected $\sim75$ galaxies ($15 \, \text{deg}^{-2} \times 5 \, \text{deg}^2$), with Poisson fluctuations resulting in $\sim77$ galaxies in our catalog (consistent with the $20 \times 20$ grid output). The catalog provides detailed positional information for each galaxy, including offsets from the field center (arcseconds/degrees) and absolute celestial coordinates (radians/degrees), stored in standard astronomical formats (FITS), forming the foundation for our informed positional prior and subsequent analysis.

\begin{figure}[htbp]
\centering
\includegraphics[width=0.48\textwidth]{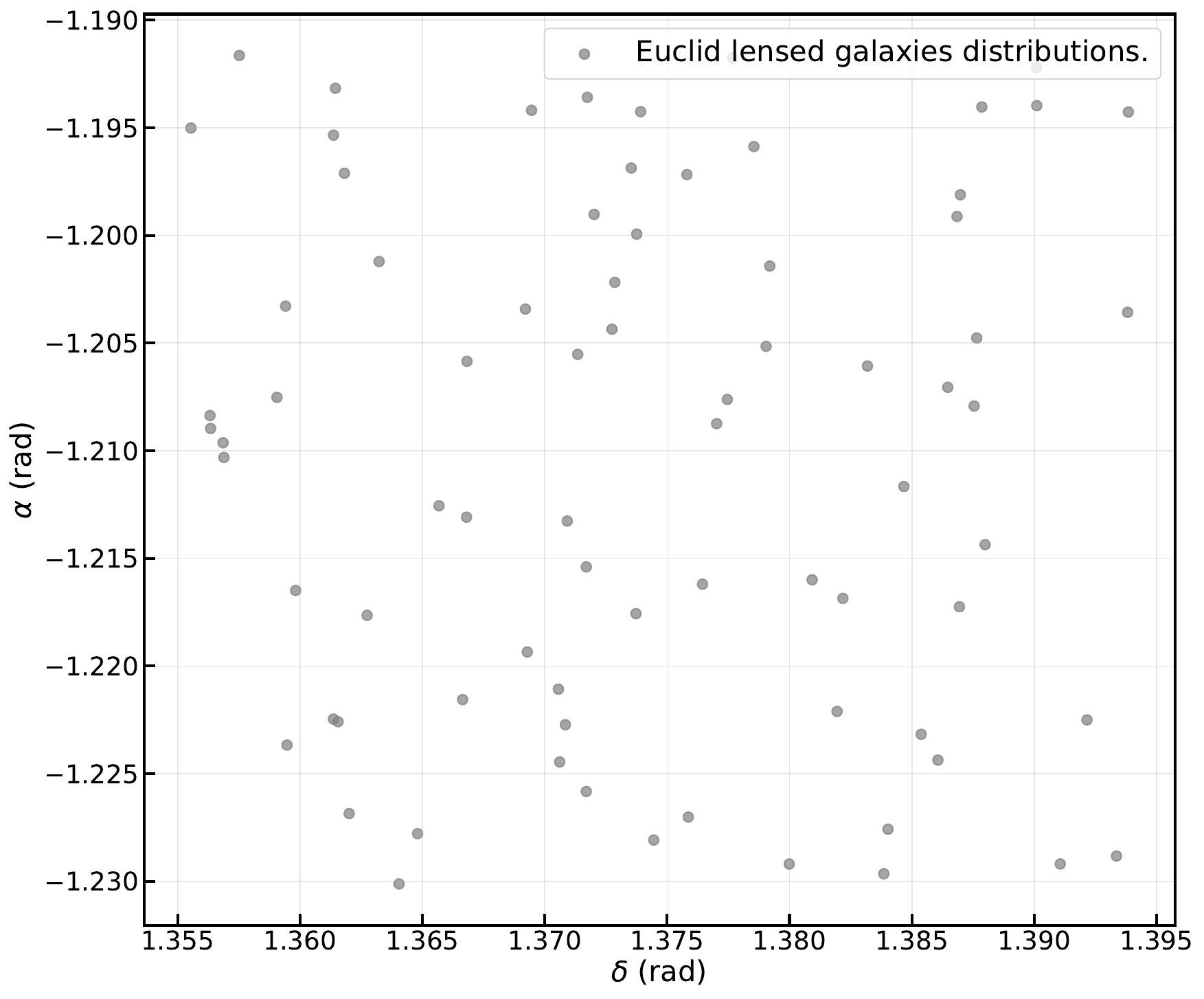}
\caption{Simulated Euclid lens galaxy catalog for informed positional priors in gravitational lensing analysis. The 5 deg$^2$ field contains $\sim77$ galaxies with a density of 15 deg$^{-2}$, matching ET's sky localization uncertainty. The field center ($\alpha=1.375$ rad, $\delta=-1.2108$ rad) corresponds to the injection position of simulated GW events.}
\label{fig1}
\end{figure}

Figure \ref{fig1} visualizes the simulated catalog, with galaxies distributed across the 5 deg$^2$ field. The gray color-coding reflects the initial spatial distribution before reweighting with GW posterior constraints, aligning with ET's sky localization uncertainty to enable robust lensing effect studies.

\subsection{Informed Positional Prior Construction}
To systematically incorporate Euclid's observational information into our GW lensing search, we construct an informative prior for the sky location of lensed events through two key steps: (1) Bayesian inference on GW data with a uniform prior to obtain posterior distributions; (2) reweighting the galaxy catalog using these posteriors to form the informed prior. This approach is grounded in the astrophysical premise that a lensed GW must originate from the same line of sight as a foreground lens galaxy, with our simulated Euclid catalog $\mathcal{G} = \{g_1, g_2, \ldots, g_N\}$ (each galaxy $g_i$ has coordinates $(\alpha_i, \delta_i)$) providing the discrete set of candidate lens positions.

\begin{figure}[htbp]
\centering
\includegraphics[width=0.49\textwidth]{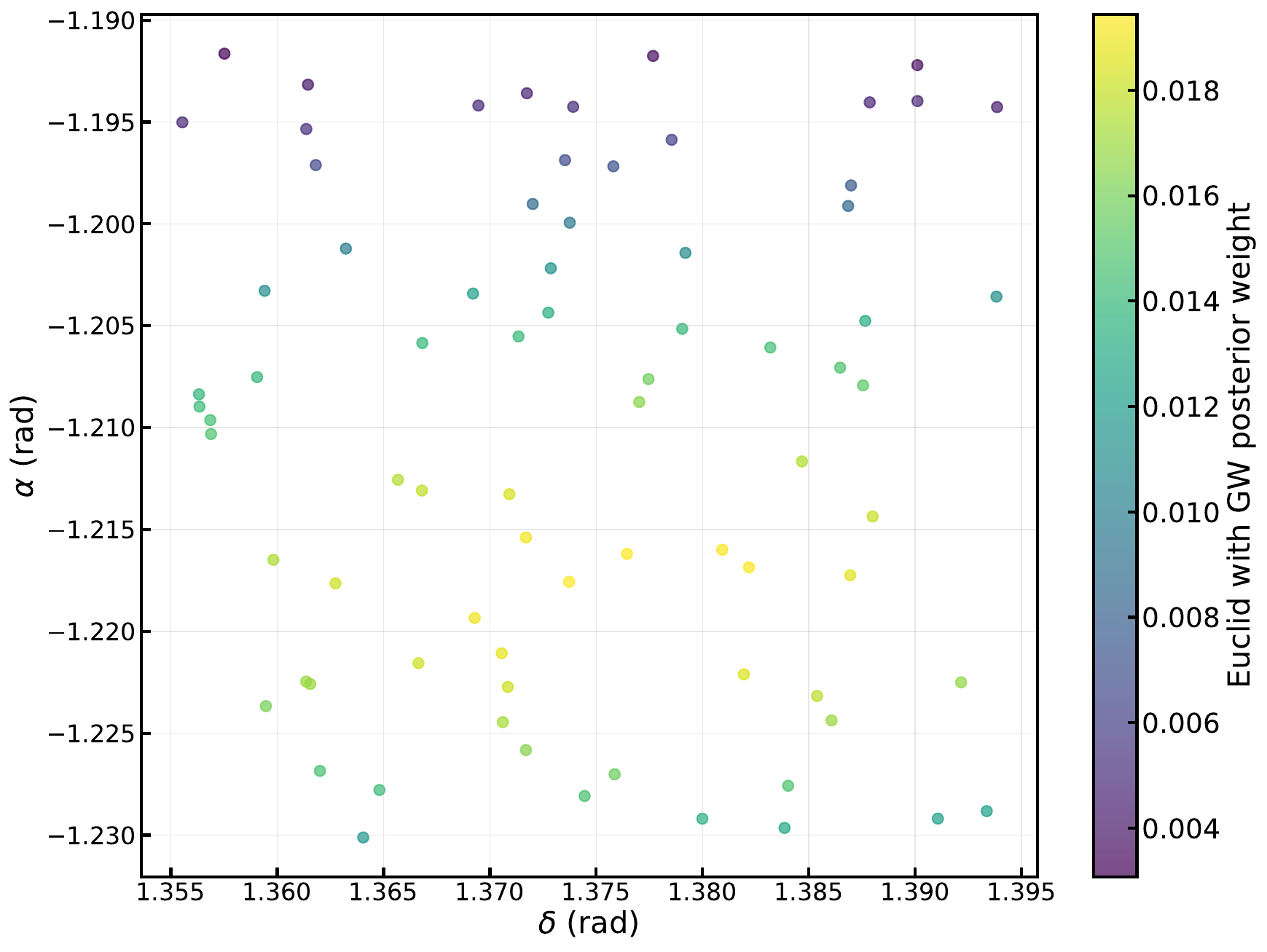}
\caption{Reweighted Euclid lens galaxy prior after incorporating gravitational wave posterior information. The initial uniform weights $w_i^{(0)} = 1/N$ have been updated to $w_i$ based on the consistency between galaxy positions $(\alpha_i, \delta_i)$ and the GW sky localization, implementing the mixture model $P_{\rm Euclid}(\alpha, \delta \, | \, \mathcal{G}) = \sum_{i=1}^{N} w_i \,K(\Delta \alpha_i, \Delta \delta_i)$. The colorbar indicates the relative probability mass assigned to each galaxy, reflecting the combined evidence from both Euclid catalog and GW parameter estimation.}
\label{fig2}
\end{figure}

We model the positional prior as a mixture of probability densities, accommodating the discrete nature of candidate lenses while accounting for continuous observational uncertainties. Initial uniform weights $w_i^{(0)} = 1/N$ are assigned to each galaxy; after obtaining the GW posterior $P_{\rm uni}(\alpha, \delta|d)$ under a uniform prior, weights are updated to reflect consistency between galaxy positions and GW sky localization:
\begin{equation}
w_i \propto w_i^{(0)} \times P_{\rm uni}(\alpha_i, \delta_i|d),
\end{equation}
with $\sum_{i=1}^N w_i = 1$ for normalization. The sky location probability density, given catalog $\mathcal{G}$, is defined as:
\begin{equation}
P_{\rm Euclid}(\alpha, \delta \, | \, \mathcal{G}) = \sum_{i=1}^{N} w_i \,K(\Delta \alpha_i, \Delta \delta_i),\label{eq:euclid_prior}
\end{equation}
where $\Delta \alpha_i = \alpha - \alpha_i$ and $\Delta \delta_i = \delta - \delta_i$. This model comprises two distinct components:
1. Discrete Weights ($w_i$): Scalar probability masses assigned to each galaxy, quantifying belief in its role as the true lens, with normalization ensuring $P_{\rm Euclid}$ is a valid probability density.
2. Spatial Smoothing Kernel ($K$): A two-dimensional isotropic Gaussian kernel that provides continuous spatial support for the discrete galaxy positions, enabling probabilistic matching with the GW localization posterior:
\begin{equation}
K(\Delta \alpha, \Delta \delta) = \frac{1}{2\pi\sigma_{\text{match}}^2} \exp\left(-\frac{(\Delta \alpha)^2 + (\Delta \delta)^2}{2\sigma_{\text{match}}^2}\right),
\end{equation}
where $\sigma_{\text{match}} = 50''$ (arcseconds) sets the characteristic angular scale for associating a GW event with a potential lens galaxy.  This scale is chosen to be conservatively larger than the intrinsic astrometric uncertainty of Euclid galaxy positions (which is $\lesssim 0.1''$), ensuring that the finite precision of the catalog does not artificially exclude potential matches. Simultaneously, $\sigma_{\text{match}}$ is significantly smaller than the typical GW sky localization area from ET (on the order of several square degrees), allowing the kernel to meaningfully discriminate between different candidate lenses within the large GW error region \citep{2022A&A...662A.112E,2025arXiv250315324E,2025A&A...697A...1E,2025arXiv250312263A}. 

Conceptually, the prior assigns discrete probability masses $w_i$ to specific galaxy coordinates, while the kernel $K$ provides a continuous probabilistic description of the event's position relative to its host lens, making it practical for Bayesian inference. Figure \ref{fig2} visualizes the reweighted prior, integrating evidence from the Euclid catalog and GW parameter estimation.

\subsection{Bayesian Inference for Lensing Identification}
The core of our lensing identification framework is the comparison between two competing hypotheses using the Bayes factor, which quantifies the evidence for a lensed origin of two GW events relative to two independent unlensed events. We first define the key hypotheses and Bayes factor formulation, then outline the two inference pipelines (uniform vs. Euclid-informed priors) and the underlying parameter estimation setup.

\subsubsection{Key Hypotheses and Bayes Factor Definition}
The problem of gravitational wave lens identification is fundamentally a comparison between two competing hypotheses:

\textbf{Lensed hypothesis} ($\mathcal{H}_{L}$): The dataset $\{d_{1},d_{2}\}$ contains lensed signals from a single binary black hole merger event with physical parameters $\bm{\theta_{1} = \theta_{2} = \theta}$ (excluding lensing-modified parameters like time delay and magnification).

\textbf{Unlensed hypothesis} ($\mathcal{H}_{U}$): The dataset $\{d_{1},d_{2}\}$ contains signals from two independent binary black hole merger events with physical parameters $\theta_{1}$ and $\theta_{2}$ being independent.

We compare these hypotheses using the Bayes factor $\mathcal{B}^{L}_{U}$, defined as the ratio of marginalized likelihoods (evidences) for the two hypotheses. Under $\mathcal{H}_{U}$, the independence of $d_1$ and $d_2$ implies:
\begin{equation}
\mathcal{Z}_{U} = P(d_{1}, d_{2} \mid \mathcal{H}_{U}) = P(d_{1})P(d_{2}),
\end{equation}
where $P(d_{i}) = \int d\theta_i P(d_i|\theta_i) P(\theta_i)$ is the marginal likelihood for event $i$. For $\mathcal{H}_{L}$, the marginalized likelihood integrates over the shared parameter space (excluding lensing-induced time delays/magnifications):
\begin{equation}
\mathcal{Z}_{L} = P(d_{1}, d_{2} \mid \mathcal{H}_{L}) = \int d\theta\;P(\theta)\;P(d_{1}|\theta)\,P(d_{2}|\theta).
\end{equation}
Using Bayes' theorem $P(\theta|d_{i}) = P(d_{i}|\theta)P(\theta)/P(d_{i})$, this rewrites as:
\begin{equation}
\mathcal{Z}_{L} = P(d_{1})P(d_{2})\int d\theta\;\frac{P(\theta|d_{1})\,P(\theta|d_{2})}{P(\theta)},
\end{equation}
yielding the compact Bayes factor:
\begin{equation}
\mathcal{B}^{L}_{U} = \frac{\mathcal{Z}_{L}}{\mathcal{Z}_{U}} = \int d\theta\;\frac{P(\theta|d_{1})\,P(\theta|d_{2})}{P(\theta)}.
\label{eq:bayes_factor}
\end{equation}

This result has an intuitive interpretation: the Bayes factor represents the inner product of the two posterior distributions, inversely weighted by the prior. If $d_{1}$ and $d_{2}$ originate from the same lensed source, their posterior distributions will show significant overlap in parameter space, favoring $\mathcal{H}_{L}$. The inverse prior weighting down-weights contributions from parameter regions with strong prior support, ensuring overlap stems from data consistency rather than prior preferences.

\subsubsection{Inference Pipelines with Different Priors}\label{priors}
We develop a systematic framework to evaluate the impact of prior choice on strong lensing identification, denoting $\mathcal{B}^{L}_{U}$ as $\mathcal{B}$ for simplicity. Our analysis implements two distinct computational pipelines, reflecting realistic data analysis scenarios with different prior assumptions.

\textbf{Uniform Prior Pipeline}: The Bayes factor is computed using posteriors from a physically motivated uniform prior:
\begin{equation}
\mathcal{B}_{\mathrm{uni}} = \int d\theta \, \frac{P_{\mathrm{uni}}(\theta|d_1) P_{\mathrm{uni}}(\theta|d_2)}{P_{\mathrm{uni}}(\theta)}.
\end{equation}
The uniform prior employs narrow ranges centered on the true injection values for all parameters, ensuring physically plausible sampling while maintaining minimal informative bias:
\begin{align}
P_{\mathrm{uni}}(\alpha) &= \mathcal{U}[\alpha_{\mathrm{true}} - 0.5\,\text{rad}, \alpha_{\mathrm{true}} + 0.5\,\text{rad}],\nonumber \\
P_{\mathrm{uni}}(\delta) &= \mathcal{U}[\delta_{\mathrm{true}} - 0.5\,\text{rad}, \delta_{\mathrm{true}} + 0.5\,\text{rad}],
\end{align}
with analogous constructions for chirp mass $\mathcal{M}_c$, mass ratio $q$, polarization angle $\psi$, spin components $a_{1}$, $a_{2}$, and inclination angle $\theta_{\rm jn}$. All prior ranges cover the 99\% credible interval of expected posteriors, ensuring no likelihood peak truncation.

\textbf{Euclid-informed Prior Pipeline}: The Bayes factor incorporates astrophysical context through galaxy catalog information:
\begin{equation}
\mathcal{B}_{\mathrm{euc}} = \int d\theta \, \frac{P_{\mathrm{euc}}(\theta|d_1) P_{\mathrm{euc}}(\theta|d_2)}{P_{\mathrm{euc}}(\theta)},
\end{equation}
where $P_{\mathrm{euc}}(\theta|d_i)$ are posteriors derived under the Euclid-informed prior.

This prior combines two components:
1. Positional parameters ($\alpha, \delta$): Kernel density estimation on the weighted galaxy catalog (see Eq.~\ref{eq:euclid_prior}), concentrating prior probability in high galaxy density regions.
2. Non-positional parameters: Identical to the uniform prior ($P_{\mathrm{euc}}(\theta_{\mathrm{non-pos}}) = P_{\mathrm{uni}}(\theta_{\mathrm{non-pos}})$), ensuring Bayes factor differences stem solely from positional information.

\subsubsection{Simulation and Parameter Estimation Setup}

To simulate lensed GW events, we generate timeseries data by injecting simulated GW signals into colored Gaussian noise consistent with the ET-D power spectral density (PSD). The signals correspond to binary black hole (BBH) mergers lensed by massive early-type galaxies, modeled with a Singular Isothermal Ellipsoid (SIE) density profile. Following astrophysical population models \cite{2018arXiv180707062H}, we simulate lensed BBH systems that produce either two (double) or four (quad) images. For quad systems, we retain the two brightest images to form a lensed pair, requiring that the fainter image in each pair has a network signal‑to‑noise ratio (SNR) $>8$, and the same threshold applied to double-image systems.

The coalescence waveforms are generated using the \texttt{IMRPhenomPv2} approximant \citep{2014PhRvL.113o1101H,2016PhRvD..93d4006H,2020PhRvD.102f4001P}, which captures the inspiral, merger and ringdown phases of BBHs with precessing spins. A reference frequency of 20 Hz is adopted, and unphysical parameter combinations are excluded through consistency checks. For lensed images, we apply a Morse phase shift of $\pi/2$ to Type II (saddle-point) images, both in doubles and in the two brightest quads \citep{2020arXiv200712709D,2023PhRvD.107l3015L}. This phase shift is not expected to significantly affect parameter estimation, as its marginalization is naturally incorporated in our likelihood construction. Each lensed image is injected into ET-like noise within the frequency band 20-1024 Hz, producing high-SNR events (typical SNRs $\sim80$) \cite{2010CQGra..27s4002P,2011CQGra..28i4013H}.

Parameter estimation is performed using the \texttt{LALInferenceNest} code \citep{2015PhRvD..91d2003V} within the \texttt{Bilby} framework \cite{2019ApJS..241...27A}, employing the \texttt{dynesty} nested sampler \citep{2019S&C....29..891H}. The likelihood is constructed from the ET‑D PSD \citep{2025arXiv250312263A}, with instrumental settings matching ET specifications: a sampling rate of 2048 Hz, 16‑second data segments (including a 2‑second pre‑signal baseline for noise estimation), and a frequency range of 20–1024 Hz. To account for realistic observational conditions, the likelihood incorporates three key marginalizations:
\begin{itemize}
\item Time marginalization: Accounts for lensing-induced time delays between lensed images and intrinsic arrival time uncertainties;
\item Phase marginalization: Handles lensing-modified Morse phase shifts for different image classes;
\item Luminosity distance marginalization: Accommodates magnification effects altering apparent luminosity distance.
\end{itemize}

Prior distributions are centered on true injection values, with ranges covering expected posterior support; details of the two types of priors considered in this work are provided in subsection \ref{priors}.
The sampler is configured with 1000 live points and parallelized over 12 cores for efficiency. Convergence is monitored using diagnostics such as the Gelman-Rubin statistic, ensuring robust posterior distributions for our lensing analysis.

\section{Results and Discussion}\label{sec3}
We present our core results in two parts: first, the robustness of intrinsic GW parameter estimation to prior choice, and second, the quantitative impact of the Euclid-informed prior on lensing identification via Bayes factor analysis. This structure isolates the effects of positional prior incorporation and validates the reliability of our key conclusions.

\subsection{Robustness of Intrinsic Parameter Estimation}
\begin{figure*}[htbp]
\centering
\includegraphics[width=0.9\textwidth]{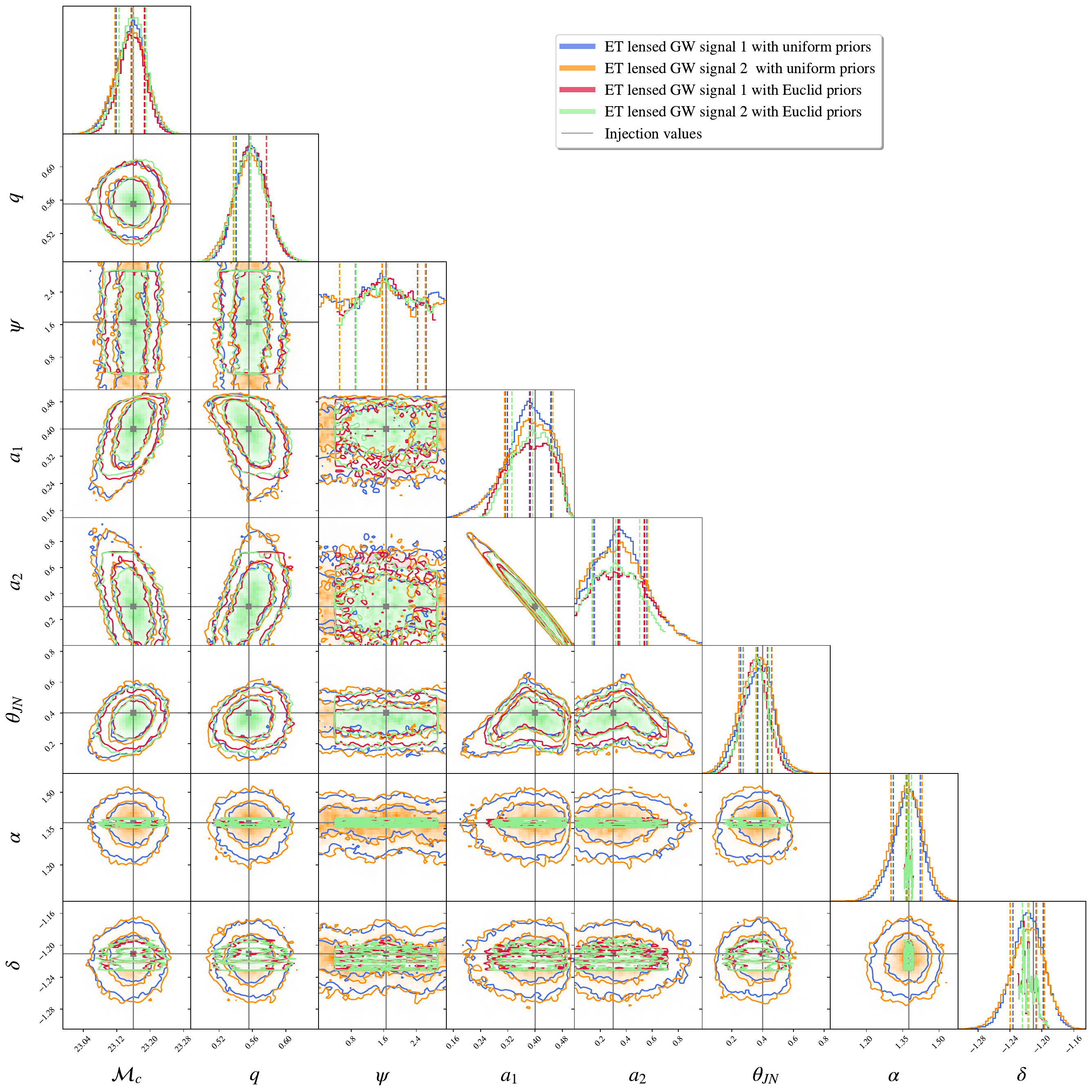}
\caption{Combined corner plot comparing posterior distributions for lensed gravitational wave signal parameters under different prior assumptions. The royal blue contours show ET lensed GW signal 1 with uniform priors, dark orange contours show ET lensed GW signal 2 with uniform priors, crimson contours show ET lensed GW signal 1 with Euclid priors, and light green contours show ET lensed GW signal 2 with Euclid priors. All contours represent the 68\% and 95\% credible regions. The injection values are marked by colored lines corresponding to each analysis (gray for uniform prior analyses, crimson and light green for Euclid prior analyses), illustrating the impact of prior information on parameter estimation accuracy. The near-perfect overlap of contours confirms that intrinsic parameter estimates are robust to prior choice for high-SNR ET signals.}
\label{fig3}
\end{figure*}

Posterior distributions of intrinsic physical parameters (chirp mass $\mathcal{M}_c$, mass ratio $q$, polarization angle $\psi$, inclination angle $\theta_{\rm JN}$, spin parameters $a_1$ and $a_2$) exhibit strong robustness to sky location prior choice (Figure~\ref{fig3}). This stems from the sharp, isolated likelihood peak of high-SNR ($\sim80$) ET signals: when the prior does not exclude the high-likelihood region, the posterior is dominated by the likelihood (Bayes' theorem: $P(\theta|d) \propto P(d|\theta)P(\theta)$), making estimates insensitive to mild prior variations.

Lensing-modified parameters (luminosity distance $D_L$, phase $\phi$, geocent time $t_c$) are marginalized over using Bilby's built-in functionality\footnote{\url{https://git.ligo.org/lscsoft/lalsuite}}, avoiding biases from magnification, phase shifts, and time delays while focusing on intrinsic source parameters. The polarization angle $\psi$ is excluded from detailed comparison due to poor constraints (typical 90\% credible interval $>180^\circ$ for ET signals).

Due to the strong dependence of Bayes factors on prior choices, direct comparison of $\mathcal{B}_{\mathrm{uni}}$ and $\mathcal{B}_{\mathrm{euc}}$ over the full parameter space would be confounded by prior differences. To isolate the effect of Euclid positional information, we first verify that intrinsic parameter posteriors are not biased by the sky location prior—a prerequisite for attributing Bayes factor differences to the improved positional prior rather than spurious parameter shifts.

Figure~\ref{fig3} compares posteriors for five well-constrained intrinsic parameters ($\mathcal{M}_c$, $q$, $\theta_{\rm JN}$, $a_1$, $a_2$) under uniform (royal blue/dark orange) and Euclid-informed (crimson/light green) priors, with injection values marked by colored lines. Contours (68\% and 95\% credible regions) show near-perfect overlap, confirming consistent means and credible intervals centered on true values.

This robustness is critical: for common intrinsic parameters $\theta_{\rm com}$, we have $P_{\mathrm{uni}}(\theta_{\rm com}|d) \approx P_{\mathrm{euc}}(\theta_{\rm com}|d)$ and $P_{\mathrm{uni}}(\theta_{\rm com}) = P_{\mathrm{euc}}(\theta_{\rm com})$ (by construction), enabling a "clean" comparison via restricted Bayes factors:
\begin{align}
\mathcal{B}_{\mathrm{uni}}^{(5)} &= \int d\theta_{\rm com} \, \frac{P_{\mathrm{uni}}(\theta_{\rm com}|d_1) P_{\mathrm{uni}}(\theta_{\rm com}|d_2)}{P_{\mathrm{uni}}(\theta_{\rm com})}, \nonumber\\ 
\mathcal{B}_{\mathrm{euc}}^{(5)} &= \int d\theta_{\rm com} \, \frac{P_{\mathrm{euc}}(\theta_{\rm com}|d_1) P_{\mathrm{euc}}(\theta_{\rm com}|d_2)}{P_{\mathrm{euc}}(\theta_{\rm com})}.
\end{align}
Any difference between $\mathcal{B}_{\mathrm{uni}}^{(5)}$ and $\mathcal{B}_{\mathrm{euc}}^{(5)}$ is unequivocally attributed to the reweighted sky location prior, laying a solid foundation for subsequent analysis.

\subsection{Bayes Factor Analysis for Lensing Identification}
The primary goal of our analysis is to quantitatively assess the enhancement in lensing identification achieved by incorporating positional information from the Euclid galaxy catalog. We focus on the Bayes factor computed using only the subset of intrinsic parameters that have identical priors in both analysis pipelines (chirp mass $\mathcal{M}_c$, mass ratio $q$, inclination angle $\theta_{\rm JN}$, dimensionless spin magnitudes $a_1$ and $a_2$), excluding lensing-modified parameters (luminosity distance, phase, geocent time) and poorly constrained parameters (polarization angle $\psi$). For these parameters, $P_{\mathrm{uni}}(\theta) = P_{\mathrm{euc}}(\theta)$, ensuring that any difference in the resulting Bayes factors $\mathcal{B}^{(5)}$ stems directly from our method of reweighting the galaxy catalog using GW posteriors.

\begin{figure}[htbp]
\centering
\includegraphics[width=0.45\textwidth]{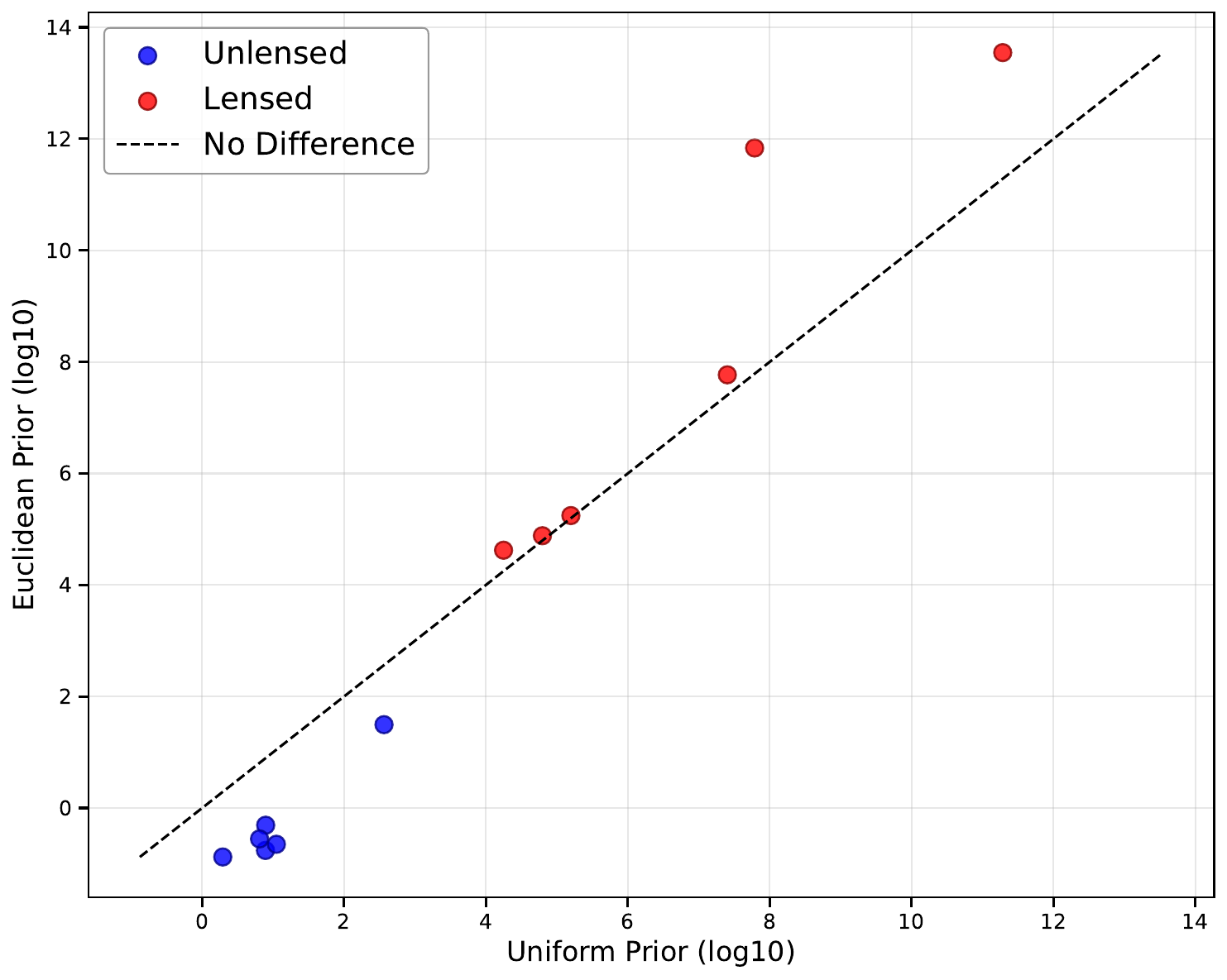}
\caption{Comparison of the restricted Bayes factors $\log_{10}\mathcal{B}^{(5)}$ computed under uniform and Euclid-informed priors. The Bayes factors are computed using only the five common parameters ($\mathcal{M}_c$, $q$, $\theta_{\rm JN}$, $a_1$, $a_2$) to isolate the effect of incorporating positional information. Blue points represent unlensed event pairs, while red points represent lensed event pairs. The dashed diagonal line indicates equality between the two priors. The systematic shift of lensed pairs above the diagonal demonstrates the enhanced sensitivity provided by the Euclid-informed prior, while the shift of unlensed pairs below the diagonal shows suppression of false positives.}
\label{fig4}
\end{figure}

Figure~\ref{fig4} presents the comparison of $\log_{10}\mathcal{B}^{(5)}$ for 12 simulated event pairs (6 lensed and 6 unlensed). A clear separation is evident: all lensed event pairs yield high Bayes factors ($\log_{10}\mathcal{B}^{(5)} > 4$) under both priors, while unlensed pairs are clustered at significantly lower values ($\log_{10}\mathcal{B}^{(5)} < 2$). Crucially, the majority of red points (lensed pairs) lie above the diagonal line, indicating that the Euclid-informed prior consistently produces stronger evidence for lensing compared to the uniform prior. Conversely, most blue points (unlensed pairs) lie at or below the diagonal, showing that the informed prior effectively suppresses false positives by down-weighting coincidental parameter overlaps that occur away from known lens galaxy positions.

The quantitative improvement is summarized in Table~\ref{tab:bf_summary}. For the six lensed event pairs, the mean Bayes factor increases from $\langle \log_{10}\mathcal{B}^{(5)}_{\text{uni}} \rangle = 7.12 \pm 1.84$ to $\langle \log_{10}\mathcal{B}^{(5)}_{\text{euc}} \rangle = 8.14 \pm 1.73$ when using the Euclid-informed prior. This corresponds to an average enhancement by a factor of $\sim 10.5$ in linear scale ($10^{8.14-7.12} \approx 10.5$). For the six unlensed pairs, the mean Bayes factor decreases from $1.98 \pm 1.25$ to $0.65 \pm 1.24$, representing a suppression by a factor of $\sim 21.4$ ($10^{1.98-0.65} \approx 21.4$). This dual effect—boosting evidence for true lensed events while suppressing it for unlensed coincidences—significantly improves the overall discriminative power of the lensing search.

\begin{table}[htbp]
\setlength{\tabcolsep}{6pt}
 \renewcommand{\arraystretch}{1.4}
\centering
\caption{Statistical summary of the restricted Bayes factors $\mathcal{B}^{(5)}$ for lensed and unlensed event pairs. The comparison uses only parameters with identical priors ($\mathcal{M}_c$, $q$, $\theta_{\rm JN}$, $a_1$, $a_2$). Uncertainties represent the standard deviation across the six event pairs in each category.}
\label{tab:bf_summary}
\begin{tabular}{lccc}
\toprule
Category & N pairs & $\langle \log_{10}\mathcal{B}^{(5)}_{\text{uni}} \rangle$ & $\langle \log_{10}\mathcal{B}^{(5)}_{\text{euc}} \rangle$ \\
\midrule
Lensed events & 6 & $7.12 \pm 1.84$ & $8.14 \pm 1.73$ \\
Unlensed events & 6 & $1.98 \pm 1.25$ & $0.65 \pm 1.24$ \\
\bottomrule
\end{tabular}
\end{table}

The physical interpretation of this result is straightforward. The reweighting procedure updates the prior probability $w_i$ for each galaxy in the Euclid catalog based on its positional consistency with the GW sky localization posterior $P_{\rm uni}(\alpha, \delta|d)$. Galaxies located within the high-probability region of the GW sky map receive increased weight, effectively concentrating the prior probability mass in astrophysically plausible locations. When computing the Bayes factor for a pair of lensed signals, this concentrated prior increases the overlap integral in regions supported by both the data and the electromagnetic catalog, leading to a higher evidence value. For unlensed pairs, whose true sky positions are independent, the reweighted prior does not favor any specific coincident location, thereby reducing the spurious overlap that might arise by chance under a uniform sky prior.

Our findings underscore a key principle in Bayesian model selection: while parameter estimation for well-measured quantities is robust to prior choice, the evidence for model comparison inherently incorporates prior information. The significant improvement in $\mathcal{B}^{(5)}$ demonstrates that the informative prior derived from the Euclid catalog provides valuable astrophysical context that is not contained in the GW data alone. This multi-messenger approach will be particularly powerful for third-generation detectors like the Einstein Telescope, where the high number of detectable lensed events ($\sim 100$ per year \citep{2022MNRAS.509.3772Y,2025arXiv251002061G}) will allow for statistical population studies. By reducing false positives and strengthening true detections, our method increases the confidence and scientific yield of strong lensing searches with gravitational waves.

A potential limitation of our approach is the reliance on a simulated Euclid catalog; future work will incorporate real Euclid data once it becomes available. Additionally, we have not yet included lens mass models or time delay priors, which could further enhance the Bayes factor by incorporating additional astrophysical constraints. These extensions will be the focus of subsequent studies.

\section{Conclusion}\label{sec4}
We have developed and demonstrated a novel Bayesian framework for identifying strongly lensed gravitational wave events by incorporating positional priors from the Euclid galaxy lens catalog. Our method features a two-step reweighting scheme: first, performing standard parameter estimation on GW data under a uniform prior; second, using the resulting posterior to reweight the galaxies in the Euclid catalog, constructing an informed positional prior that reflects the consistency between potential lens locations and the GW sky localization.

Through a controlled comparison of Bayes factors computed over the subset of intrinsic parameters with identical priors, we show that the Euclid-informed prior enhances the Bayes factor for truly lensed event pairs by an average factor of $\sim 10.5$ while suppressing it for unlensed pairs by a factor of $\sim 21.4$. This dual effect—strengthening true detections and reducing false positives—significantly improves the discriminative power of lensing searches. Importantly, we verify that intrinsic GW parameter estimates remain robust to prior choice, ensuring that the improvement in lensing identification stems purely from the astrophysical positional information provided by Euclid.

This work highlights the critical importance of prior specification in gravitational wave lensing searches, especially for model selection tasks. It provides a practical methodology for integrating electromagnetic survey data into GW analyses, paving the way for more confident detections and population studies of strongly lensed gravitational waves with next-generation observatories. Future extensions of this framework will include lens mass models, time delay priors, and real Euclid data, further enhancing the sensitivity and physical interpretability of lensing searches.

\section*{Acknowledgments}
This work was supported by National Key R$\&$D Program of China (No. 2024YFC2207400), the National Natural Science Foundation of China under Grants No. 12203009. 
\bibliography{ref}

\end{document}